\documentstyle[12pt,aasms4]{article}
\input epsf
 
\def\k{km s$^{-1}$}
\def\pp{^{\prime\prime}}
\def\cm2{cm$^{-2}$}
\def\c3{cm$^{-3}$}

\begin{document}

\title{CO observations toward supernova remnants with associated OH 1720 MHz 
masers} 
 
\author{Estela M. Reynoso, \altaffilmark{1}} 
\affil{Instituto de Astronom\'\i a y F\'\i sica del Espacio,\\ CC 67  Suc 28,
(1428) Buenos Aires, Argentina}
\author{and Jeffrey G. Mangum}
\affil{National Radio Astronomy Observatory, 949 North Cherry Avenue, Tucson, 
AZ 85721-0655, USA}
 
\altaffiltext{1}{Member of the Carrera del Investigador Cient\'\i fico, 
CONICET, Argentina} 
\authoremail{ereynoso@iafe.uba.ar}
 
\begin{abstract}

The environs of three supernova remnants (SNR)  with associated OH 1720 MHz
masers, G349.7+0.2, CTB 37A and G16.7+0.1, have been surveyed in the CO J=1-0
transition with the 12 Meter Telescope of the NRAO, using the On-The-Fly 
technique. These observations have revealed a number of molecular clouds
interacting with the SNR shock fronts. Most of the OH 1720 MHz masers have
been found to lie over CO concentrations, and the maser velocities are 
coincident with the CO peak velocities to an accuracy better than 2 km/s. The 
present data trace the interstellar medium (ISM) structures interacting with 
the SNRs; however, to probe the shocked molecular gas in which the OH 1720 MHz 
emission originates, higher excitation transitions and more complex species 
should be observed. In CTB 37A, where the shock velocity into the molecular 
cloud could be determined, it has been found to be of C-type, in agreement with 
theoretical predictions. Part of the rim of G16.7+0.1 appears to be flattened 
by a dense external cloud, yet the only associated OH 1720 MHz maser lies near 
the opposite region of the remnant. This behavior, also observed in IC 443 and 
3C 391, seems to contradict the suggestion that OH 1720 MHz maser emission 
occurs mainly for transverse shocks. 

\end{abstract}

 
\section {Introduction:}

When a SNR expands near or inside a molecular cloud, the 
shock driven into the cloud can accelerate relativistic particles, heat and 
compress the molecular gas, and change its chemistry, in addition to being an 
important source of turbulent mixing. If the turbulence is not strong enough as
to disrupt the cloud, condensed clumps may be created which eventually may 
become new stars. Therefore, the interaction of SNRs with molecular clouds 
joins the two endpoints of the life cycle of stars and gas in our Galaxy 
(Frail, Goss \& Slysh 1994). The study of such interactions sheds light on 
several different fields in astrophysics. 

Unfortunately, confusion introduced by unrelated gas along the line of sight,
makes it not trivial to establish unambiguously whether an SNR is physically 
associated with a molecular cloud or whether their distances differ and they 
appear as positionally coincident just because of a projection effect. 
Morphological signatures like arcs of gas outlining an SNR, or indentations in 
SNR edges surrounding gas concentrations, are often used (Landecker et al. 
1989, Reynoso et al. 1995, Dubner et al. 1999). More convincing, but also more 
rare, are line broadenings (e.g. Frail \& Mitchell 1998, Reach \& Rho 1999) or 
wings (Seta et al. 1998) indicative of shocked gas. 
 
Almost three decades ago, Goss (1968) discovered for the first time OH emission 
in the satellite line at 1720 MHz associated with two SNRs: W28 and W44 (see
also Goss \& Robinson 1968). Hardebeck (1971) found that this emission comes
from several compact regions with strong inversion of the OH levels. Later, 
another OH 1720 MHz 
emission feature was found towards IC 443 (DeNoyer 1979a). Turner (1969) 
suggested that the inversion of the 1720 MHz line is produced by infrared 
pumping. Since then, however, galactic OH maser emission at 1720 MHz near SNRs 
has been largely forgotten. In the following years, the evidence that the three 
SNRs above are interacting with nearby clouds has grown as to make them three 
of the best known and studied cases of SNR-molecular cloud interactions (eg. 
Wootten 1977, 1981, DeNoyer, L.K. 1979b, 1983, White et al. 1987, Burton et al. 
1988). It is natural then to suggest that this interaction plays a key role in 
the production of maser emission at 1720 MHz (Frail et al. 1996). Several OH 
1720 MHz surveys have been carried out recently (Frail et al. 1996, Green et 
al. 1997, Koralesky et al. 1998) in order to find new cases of SNR-OH 1720 MHz 
maser associations. At present, among 150 surveyed Galactic SNRs, positive 
results have been obtained for around twenty of them.  

The origin of this maser emission can be easily explained through collisional
excitation by the passage of a shock front (Frail et al. 1996). Therefore, the 
presence of OH 1720 MHz masers is proposed as a powerful tool to diagnose 
SNR-molecular cloud interactions. Elitzur (1976) showed that the pumping is
more efficient if the kinetic temperature is in the range 25 to 200 K, and the
H$_2$ density is in the range $10^3$ to $10^5$ \c3. Locket, Gauthier \& Elitzur 
(1999) found tighter constrains on the physical conditions needed to produce OH 
1720 MHz maser emission: the presence of masers imply moderate temperatures (50 
-- 125 K) and densities of the order of 10$^5$ \c3, which can exist only if the 
shocks are of C-type.  Thus, the abscence of
OH 1720 MHz masers in itself is not enough to rule out an interaction. Since 
the maximum amplification occurs along the edge of an SNR, where the least
velocity dispersion and hence the largest coherence is maintained, the velocity 
of an OH 1720 MHz maser associated to an SNR can be identified with the 
systemic velocity of the remnant, making it possible to constrain its distance. 
This property is also suggested by Claussen et al. (1997), who point out 
that the only OH 1720 MHz maser emission observed toward IC 443 appears at the 
site where clear evidence for the existence of a transverse shock is found.

In a recent paper, Frail \& Mitchell (1998) surveyed the regions of the masers 
associated with the well studied SNRs W28, W44 and 3C 391, in high excitation 
CO and H$_2$CO transitions. The latter SNR is also known to be interacting with 
molecular gas (Wilner, Reynolds \& Moffett  1998, Reach \& Rho 1988, 1999). The 
authors succeeded in finding the molecular shocks in all three of them, 
supporting the idea that the masers originate in post shock gas. They found 
that the masers are distributed near, but not coincident with, peaks of CO 
J=3--2 emission.

Motivated by the proposed interaction between SNRs and molecular clouds
that OH 1720 MHz masers may reveal, we have selected three SNRs: G349.7+0.2,
G16.7+0.1 and CTB 37A, and surveyed the molecular gas around them in search
for the clouds hypothetically interacting with the shock fronts. We used the 
lowest energy transition of the CO molecule as a probe, which is ubiquitous and 
easy to detect, as a first approach to the structure of the ISM around these 
remnants. The present observations allowed us to detect several sites of 
possible interaction between these SNRs and the surrounding molecular gas, even 
at locations where no OH 1720 MHz masers are present (e.g. G16.7+0.1). These 
observations provide the basis for future work analyzing molecular shocks 
through higher transitions, and for comparing the physics of shocked and 
unshocked gas. 

\medskip
\section {Observations and data reduction:}

Observations of the CO J=1--0 line (115.271204 GHz) were performed with the
NRAO\altaffilmark{2} \altaffiltext{2}{The National Radio Astronomy Observatory 
is a facility of the National Science Foundation operated under cooperative 
agreement by Associated Universities, Inc.} 12 Meter Telescope located on Kitt 
Peak, Arizona, on 1998 June 12-18 and 27-30. The beam size at this frequency is 
$54\pp$. A square field of $\sim 38^\prime \times 38^\prime$ was covered toward 
each of the remnants. Observations were made using the On-The-Fly (OTF) 
technique, allowing efficient imaging of extended regions. Three difrerent 
spectrometers were employed: two of them with 128 channels and resolutions of 1 
MHz and 500 kHz, respectively, and a 768-channel correlator with a resolution 
of 98 kHz. An absolute position-switching mode was used, with reference 
positions given in Table {\ref{tbl1}}. These reference positions were 
measured to be free of CO emission down to levels of less than about 0.5 K.
Four images were obtained for G349.7+0.2, five for G16.7+0.1 and two for 
CTB 37A. Since data at the 12 Meter Telescope are calibrated by default to 
the T$^*_R$ scale, which is equivalent to the radiation temperature T$_R$ for 
a source much larger than the main diffraction beam of the telescope, it
is often necessary to apply corrections which refer to sources of
different sizes.  For sources with sizes in the range $20\pp -40\pp$, the
efficiency factor which converts T$^*_R$ to T$_R$ is 0.85 based on historic
measurements of the planets Jupiter and Saturn.

The AIPS package was employed for data processing. Baselines were subtracted
in the UV plane directly with the task SDLSF, which uses a unique range of
line-free channels in the whole area. Because of the proximity of all three 
sources to the Galactic plane, the intensity distribution along the observed 
images made it extremely difficult to find a suitable range of line-free 
channels. We thus decided to give priority to the baseline removal in the 
regions closest to the remnants and OH 1720 MHz masers, with possible detriment 
to the rest. Images were averaged for each source using the task WTSUM, with 
$1/\sigma^2$ as weight, where $\sigma$ is the noise image corresponding to each 
image. Due to the very high noise of one of the original images of G349.7+0.2, 
only three of them were kept in the average for this source.

In Table {\ref{tbl1}}, we summarize the observational parameters, including the 
noise of the final images for each SNR.

\placetable{tbl1}

\begin{deluxetable}{lccc}

\tablecaption{Observational Parameters\label{tbl1}}
\startdata
&\nl
\tableline
\tableline
Frequency [GHz]:&115.27 (2.6 mm)\nl
Spectral resolution [kHz, \k]&98 &0.25 &768\nl
and number of channels:&500&2.6 &128\nl
&1000&1.3&128\nl
Angular resolution [arcsec]:&53.6$\times$53.6\nl
Surveyed area [arcmin]:& $\sim 38 \times 38$\nl
Observing mode :& absolute position switching\nl
\tableline
&G349.7+0.2&CTB37A&G16.7+0.1\nl
Central position [RA (1950)]& 17 14 36.0&17 11 00.0&18 18 07.4 \nl
{\hskip 3 cm} [decl. (1950)]:& -37 22 59&-38 30 00&-14 23 18\nl
Reference position [RA (1950)]& 17 00 36.0&16 35 00.0&18 35 07.4 \nl
{\hskip 3 cm} [decl. (1950)]:& -37 22 59&-38 00 00&-14 23 18\nl
Number of images:&3&2&5\nl
Observing time [hours]:&4.1&3.6&8.3\nl
Central velocity (LSR)  [\k ]:&--40&--65&+50\nl
rms [K]:&0.6 &0.8 &0.3 \nl
\enddata

\end{deluxetable}

\medskip
\section {Results:}

In what follows, we will devote a subsection to each SNR, including a
brief introduction to the remnant, a description of our CO results and an
estimation of physical parameters for the structures most probably related 
to the remnant. Distances shall be derived from the Galactic rotation model
by Fich, Blitz \& Stark (1989). To calculate the H$_2$ column density, the
relationship $X=N(H_2)/W_{CO}$ will be used, where $W_{CO}$ is the integrated 
CO line intensity, $W_{CO}=\int{T_R(CO)}\, dv$, in K \k , and where we have 
divided our antenna-based T$^*_R$ temperatures by the efficiency factor of 
0.85 to place them on the T$_R$ scale appropriate for sources with sizes in the 
range $20\pp-40\pp$. Several studies suggest a trend of increasing $X$ with 
Galactocentric distance (Digel et al. 1997), thus the value of $X$ may vary 
from one remnant to another.

Masses will be estimated by integrating $N_{H_2}$ over the solid angle subtended
by the CO emission feature. Assuming spherical volumes and using a mean 
molecular weight per H$_2$ molecule of 2.72 m$_H$ (Allen 1973), the mass and 
density of a cloud can be expressed as:
\begin{equation}M=5.65\times 10^{-21}\, X\, W_{CO}\, \theta^2\, D^2\ \ M_
\odot\end{equation}
\noindent and
\begin{equation}n_{H_2}={{4.1\times 10^{-19}\, X\, W_{CO}\, \sqrt f}\over{D\, 
\theta}}\ \ {\rm cm}^{-3},\end{equation}
where $D$ is the distance in kpc, $\theta$ is the angular radius in arcmin of 
(the cloud assumed spherical), computed as half the measured diameter 
deconvolved by the beam size, and $f$ is a correction factor allowing for 
elongated shapes, and is given by the ratio $\theta_{maj}/\theta_{min}$.  

\medskip
\subsection {G349.7+0.2:}

This remnant consists of an incomplete clumpy shell, $2^\prime$ in diameter,
with enhancement to the south (Figure 1). HI observations (Caswell et al. 
1975) suggest that G349.7+0.2 is located beyond the tangent point, 
since absorption features are detected even at positive velocities, 
up to about +6 \k . Clark \& Caswell (1976) notice that the strong north-south 
emission ridge, somewhat east of the center, probably represents a 
non-uniform distribution of emission in a shell  rather than a filled
center, and would be evidence for a strong interaction with the ISM at an 
early stage in the life of this SNR. Frail et al. (1996) detected five OH 
1720 MHz masers, with the brightest of them aligned along the emission ridge, 
and a few weak ones at the southern continuum knot.

Our CO observations clearly reveal a concentration inside the remnant, at 
RA=$17^h 18^m 0^s.0$, decl=$-37^\circ 26^\prime 36\pp$ (J2000). This
concentration appears just above the southern knot, coincident with the 
emission ridge. A gaussian fit  to the spectrum of this cloud yields a central 
velocity of +16.5 \k \ and a FWHM velocity of $\Delta v=4$ \k . Figure 2 shows 
the cloud integrated over this $\Delta v$. There is a striking coincidence 
among the CO cloud and the OH 1720 MHz masers, which appear encircling the 
former. From Table 2 in Frail et al. (1996), the velocities of the OH 1720 MHz 
masers vary from +14.3 to +16.9 \k , in excellent agreement with the velocity 
of the CO cloud.

If we adopt the systemic velocity of the cloud to be +16.5 \k , the distance 
turns out to be $\sim 23$ kpc, in close agreement with previous results (Frail 
et al. 1996). The diameter of the cloud is estimated to be $\sim 50$ arcsec, 
which corresponds to 5 pc. This CO cloud is placed at a Galactocentric distance 
of $R=14$ kpc. For $R$ = 12-13 kpc, Digel, Bally \& Thaddeus (1990) estimate 
that $X=(8\pm 4) \times 10^{20}$ \cm2 (K \k )$^{-1}$, while Mead \& Kutner 
(1988) give a value of $\sim 4$ \cm2 (K \k )$^{-1}$. Therefore, we use 
$X=8\times 10^{20}$ \cm2 (K \k )$^{-1}$ keeping in mind that the mass and 
density may be over estimated by a factor of 2. With these values, the mass and 
H$_2$ number density yield 1.2$\times 10^4$ M$_\odot$ and 1100 \c3, 
respectively.

In Table  {\ref{tbl2}} we summarize the results obtained above. The second 
column quotes the approximate central position of the structure in equatorial 
coordinates referred to J2000. The third column lists the radius $r$ after 
deconvolution with the beam size. The fourth and fifth columns contain the 
systemic velocity $v_{sys}$ and the FWHM velocity $\Delta v$, both in \k. The 
computed mass and H$_2$ density are given in the last two columns. 

\placetable{tbl2}

\begin{deluxetable}{lcccccc}

\tablecaption{Observed and derived parameters for the CO structure
associated with G349.7+0.2\label{tbl2}}
\tablehead{
        \colhead{central coordinates} &
        \colhead{r } &
        \colhead{$v_{sys}$}&
	\colhead{$\Delta v$}&
	\colhead{M} &
	\colhead{$n_{H_2}$}\nl
        \colhead{} &
        \colhead{RA, decl. (J2000)} &
        \colhead{(arcsec)} &
        \colhead{(\k)}&
	\colhead{(\k)}&
	\colhead{(M$_\odot$)} &
	\colhead{(\c3)}}
\startdata
17 18 0.0, -37 26 36&24&+16.5&4.0&1.2$\times 10^4$&1100\nl
\enddata

\end{deluxetable}

\medskip
\subsection {CTB 37A:}

This source (Fig. 3) is actually two SNRs overlapped in projection (Kassim,
Baum \& Weiler 1991). One of them, G348.5+0.1 (to the west in Fig. 3), is a 
partial shell with a faint extension out the open end of the shell (``breakout''
morphology), while the other one, G348.5-0.0 (to the east in Fig. 3), taken to 
be a jet of the former by Milne et al. (1979), is a partial shell. By comparing 
HI spectra of CTB 37A with G349.7+0.2, which is about one degree apart, Caswell 
et al. (1975) conclude that CTB 37A is beyond the tangential point (at --110 
\k ) but not farther away than a feature seen at --65 \k .  While most of the 
OH 1720 MHz masers identified by Frail et al. (1996) appear projected over 
G348.5+0.1 between --63.5 and --66.3 \k , two of them fall at --21.4 and 
--23.3 \k \ respectively. These two latter masers appear in the region where 
the two remnants overlap, suggesting that they are associated with G348.5-0.0.
Therefore, the OH 1720 MHz masers allow an assignement of the systemic 
velocities of the two remnants that constitute CTB 37A.

An inspection of the CO emission toward CTB 37A throughout the whole
observed velocity range, reveals a very complex distribution. In searching
for a positional coincidence among the OH 1720 MHz masers and the CO emission, 
the best association could be found from --22.7 to --25.3 \k \ (Fig. 4) and
from --60.4 to --68.3 \k \ (Fig. 5). In what follows we will describe the 
identified structures likely to be interacting with the CTB 37A complex.

At about --24 \k, a weak CO concentration is observed to the west of
G348.5-0.0, with a few brighter clumps embedded. This structure, centered
approximately at RA=$17^h 14^m 39^s$, decl=$-38^\circ 30^\prime $ (J2000), 
shall be called hereinafter ``the western cloud''. The two OH 1720 MHz masers 
at --21.4 and --23.3 \k \ lie on the edge of the cloud. A Gaussian fit to the 
average profile towards the western cloud yields a central velocity of 
--23.1 \k \ with a FWHM velocity of 5.3 \k. At this systemic velocity, the 
kinematical distance could be 3.1 or 13.5 kpc. If we assume that this cloud 
is related to the CTB 37A complex, the HI absorption study of Caswell et al. 
(1975) solves the ambiguity  in favor of the larger distance. Regardless of 
the ambiguity, the Galactocentric distance is 5 kpc. 

A value of $X=2 \times 10^{20}$ \cm2 (K \k )$^{-1}$ will be used in deriving 
mass and mean H$_2$ density, based on $\gamma-$ray emission studies for the
inner Galaxy (Lebrun et al. 1983). Assuming an ellipsoidal geometry with the 
major and minor axes equal to 340$\pp$ and 135$\pp$ respectively, the mass is 
estimated to be $\sim 15900$ M$_\odot$, and the average H$_2$ density, 150 \c3.

Other structures probably associated with the OH 1720 MHz masers are found 
beyond --60 \k . In the integrated image shown in Fig. 5, there is an 
extended structure covering the north-west half of G348.5+0.1. Several 
components can be recognized inside this large structure. Because of the 
proximity with the OH 1720 MHz masers, we will focus on two of them: one 
centered near RA=$17^h 14^m 20^s.5$, decl=$-38^\circ 33^\prime 30\pp$ (J2000) 
with three OH 1720 MHz masers on its northern edge, which shall be called ``the 
central cloud''; and another one located to the north, coincident with the 
continuum shell from RA=$17^h 14^m 50^s.5$, decl=$-38^\circ 27^\prime 10\pp$ 
(J2000) to RA=$17^h 14^m 17^s.5$, decl=$-38^\circ 30^\prime 40\pp$ (J2000), 
which shall be called ``the northern cloud''. Another small, weak cloud which
appears detached from the bulk of the CO emission will also be studied. This
cloud, hereafter called ``the southern cloud'', is centered approximately at
RA=$17^h 14^m 28^s.5$, decl=$-38^\circ 41^\prime 50\pp$ (J2000) and contains
an OH 1720 MHz maser on its northern edge.
 
Since the three clouds are at --65 $\pm 2$ \k , we adopt a Galactocentric 
distance of 3.5 kpc for them, and a kinematical distance of 11.3 kpc, where 
again we have discarded the closer distance of 5.3 kpc based on the HI 
absorption results. The central and FWHM velocities
of each cloud, as well as the derived masses and densities, are listed in
Table {\ref{tbl3}}. In this table, we also include the western cloud at --23.1 
\k . Units are as in Table {\ref{tbl2}}. In the third column, the listed radii
are derived assuming spherical geometries and, where applicable, the correction 
factor $f$ is included in parenthesis. The northern cloud is assumed to be 
an ellipsoide enclosed by the outermost white contour in Fig. 5. In fitting a 
Gaussian to the profile towards the central cloud, only the highest peak was 
taken into account, since the profile looks highly asymmetric and  consisting 
of several components. For all the clouds, $X$ was assumed to be equal to $2 
\times 10^{20}$ \cm2 (K \k )$^{-1}$. 

\placetable{tbl3}

\begin{deluxetable}{lcccccc}

\tablecaption{Observed and derived parameters for CO structures 
associated with CTB 37A\label{tbl3}}
\tablehead{
        \colhead{cloud} &
        \colhead{central coordinates} &
        \colhead{r } &
        \colhead{$v_{sys}$}&
	\colhead{$\Delta v$}&
	\colhead{M} &
	\colhead{$n_{H_2}$}\nl
        \colhead{} &
        \colhead{RA, decl. (J2000)} &
        \colhead{(arcsec)} &
        \colhead{(\k)}&
	\colhead{(\k)}&
	\colhead{(M$_\odot$)} &
	\colhead{(\c3)}}
\startdata
Western &17 14 39, -38 30&107 (2.5)&--23.1&5.3&1.5$\times 10^4$&150\nl
Central &17 14 20.5, -38 33 30&60 (2.1)&--64.0&6.9&7.2$\times 10^3$&520\nl
Northern &17 14 30,  -38 27 &150 (3.13)&--65.0&12.0&5.8$\times 10^4$&660\nl
Southern &17 14 28.5,  -38 41 50&37 &-66.6&4.0&1.3$\times 10^3$&280\nl
\enddata

\end{deluxetable}

\medskip

\subsection {G16.7+0.1:}

G16.7+0.1 (Fig. 6), which belongs to the composite type of SNRs, consists
of an irregular shell, brighter to the south, and a central jet-like feature
probably powered by an undetected pulsar (Frail \& Moffett 1993). Green et al.
(1997) detected a single OH 1720 MHz maser near the bright southern edge of the 
shell, at +20 \k . Unfortunately, there are no line absortion studies to 
constrain the systemic velocity of the remnant and compare it to the velocity
of the OH 1720 MHz maser.

Like CTB 37A, G16.7+0.1 appears to be located in a very complex region. The
CO emission is extended  over a large velocity range. The only feature
found at the position of the OH 1720 MHz maser is a  small cloud at $\sim +25$ 
\k , centered approximately at RA=$18^h 20^m 56^s.2$, decl=$-14^\circ 21^\prime 
55\pp$ (J2000). This feature will be called ``the southern cloud''. The 5 \k
\ gap between the velocities of the OH 1720 MHz maser and the southern cloud, 
may be related to the anomalous velocity dispersion of the maser, larger than
typical values (Green et al.  1997). The southern cloud is seen from +25.1 to  
+25.9 \k . Figure 7 shows the CO emission integrated over this velocity range.

There are two more structures that seem to be associated with the remnant. One 
of them is a small concentration inside the shell, centered at about RA=$18^h 
20^m 59^s.6$, decl=$-14^\circ 19^\prime 22\pp$ (J2000), located to the east of 
the central continuum feature. In what follows, we will call this concentration
``the central cloud''. The other structure is a bright concentration to the 
northwest, hereinafter ``the north-western cloud'', centered approximately at 
RA=$18^h 20^m 46^s.6$, decl=$-14^\circ 18^\prime 10\pp$ (J2000). The close 
agreement among the contours of the CO and continuum emission, strongly 
suggests that the north-western cloud is causing the flattening of the shock 
front observed in this direction. 

The parameters computed for these three clouds are listed in Table {\ref{tbl4}}.
To estimate the distance, we first notice that the peak velocities of the clouds
are within 0.2 \k~ of +25.6 \k. At +25.6 \k , the Galactic rotation model 
predicts two kinematical distances: 2.6 and 13.7 kpc, and a Galactocentric 
distance of 3.5 kpc. As stated above, there are no absorption studies to solve 
the kinematical distance ambiguity. Therefore, we attempted to find which of 
the two distances is most likely applicable using a $\Sigma$-$D$ relationship, 
in spite of the large intrinsic errors of this method (typical uncertainties 
can be more than a factor of 2). Based on the surface density given by Helfand 
et al. (1989), the $\Sigma$-$D$ curve estimated by Case \& Bhattacharya (1998) 
produces a distance of 13.7 kpc, in excellent agreement with the farther
distance given by the Galactic rotation model. Though this perfect agreement 
may a coincidence, the $\Sigma$-$D$ distance implies that the lower kinematical
distance should be discarded. The ratio $X$ is assumed to be 
equal to $2 \times 10^{20}$ \cm2 (K \k )$^{-1}$. The units are as in Table 
{\ref{tbl2}}. The third column is as described in Table {\ref{tbl3}}.

\placetable{tbl4}

\begin{deluxetable}{lcccccc}

\tablecaption{Observed and derived parameters for CO structures 
associated with G16.7+0.1\label{tbl4}}
\tablehead{
        \colhead{cloud} &
        \colhead{central coordinates} &
        \colhead{r } &
        \colhead{$v_{sys}$}&
	\colhead{$\Delta v$}&
	\colhead{M} &
	\colhead{$n_{H_2}$}\nl
        \colhead{} &
        \colhead{RA, decl. (J2000)} &
        \colhead{(arcsec)} &
        \colhead{(\k)}&
	\colhead{(\k)}&
	\colhead{(M$_\odot$)} &
	\colhead{(\c3)}}
\startdata
Southern&18 20 56.2, -14 21 55&32 (1.6)&+25.8&1.5&$2.1\times 10^3$&260\nl
Central &18 20 59.6, -14 19 22&28 (1.5)&+25.4&4.4&1.8$\times 10^3$&330\nl
North-western &18 20 46.6, -14 18 10&37 (2)&+26.6&2.7&3.4$\times 10^3$&280\nl
\enddata

\end{deluxetable}

\medskip
\section {Discussion:}
\medskip
\subsection {Correlation between CO clouds and OH 1720 MHz masers}

To compare quantitatively the agreement between the velocities of the OH 1720
MHz masers and the velocities of the CO peak flux densities, CO profiles toward 
all individual masers were taken, and Gaussian fits to the identified peaks 
were calculated. The average difference in velocities is $|\Delta V|=0.6$ \k,
with a disperssion of 1.7 \k. The highest departure, above 5 \k, corresponds to 
G16.7+0.1. Disregarding this latter value, $|\Delta V|$ turns out to be 
$0.2\pm 1.2$ \k.

In addition to the velocity agreement, the positional agreement between CO 
clouds and OH 1720 MHz masers is quite striking in many cases. The best match 
is found in G349.7+0.2, where all of the masers are aligned around the only CO 
feature detected towards the remnant. No correlation is found between maser 
peak flux densities and proximity to the CO emission peak. Brogan et al. (2000)
were able to measure magnetic fields of $\sim 300 \mu$G using the Zeeman effect
towards the brightest OH masers in this SNR. 

Another case in which a very good agreement is found is G16.7+0.1. The OH 1720
MHz maser is not only superimposed on a CO clump (the southern cloud) but also
lies at the contact region between the southern cloud and a continuum 
maximum. However, the velocities are not coincident. According to Green
et al. (1997), the maser is at $\sim +20$ \k. The CO images around this
velocity show no emission near the position of the OH 1720 MHz maser. In the 
next paragraph, some suitable explanations for this discrepancy are 
proposed. 

Claussen et al. (1997) found that some of the OH 1720 MHz masers around the SNR 
W28 do not share the systemic velocity of the remnant, but are blueshifted by
$\sim 10$ \k, as is the molecular gas associated with them. The CO profile 
towards the southern cloud in G16.7+0.1, shows a weak wing centered at +19.3 
\k , with a 6.5 \k \ FWHM. This wing, which is in the limit of detectability,
would imply that part of the molecular gas of the southern cloud was 
accelerated to $\sim 6.5$ \k \ by the shock wave and, like in the case of W28, 
the OH 1720 MHz maser indicate the velocity of the accelerated gas. It is also 
possible that the maser is in fact blended features, as Green et al. (1997) 
suggested. In such case, it would be important to compare the velocities of 
splitted components to the systemic velocity of the southern cloud to analyze 
the agreement among them.

In CTB 37A, the correlation is poor compared to the other two SNRs. Although
the masers are in most cases located inside CO structures, the proximity to
emission peaks does not always apply. Moreover, the strength of the masers is 
unrelated to the local CO emission intensity. There are two masers, located at 
RA=$17^h 13^m 54^s.87$, decl=$-38^\circ 33^\prime 25\pp.5$ (J2000) and RA=$17^h 
13^m 58^s.81$, decl=$-38^\circ 34^\prime 26\pp.2$ (J2000), with velocities
--65.1 and --65.2 \k \ respectively, which lie outside the edge of the CO 
emission region. Brogan et al. (2000) found that, unlike W28 and W44 where
magnetic fields are uniform both in magnitude and direction, in CTB 37A the
magnetic field strength $B_\theta$ changes direction along the line of sight 
over length scales of only $\sim 3$ pc, revealing a complicated magnetic field
morphology.

The CO features found with the present data to be associated with the OH 1720 
MHz masers, have densities not higher than $\sim 10^3$ \c3, which is at least 
two orders of magnitude lower than theoretical predictions. 
The $^{12}$CO J=1--0 line is highly saturated toward the dense center of 
molecular clouds, and thus does not vary linearly with the column density 
(Oka et al. 1998). However, low-density cloud envelopes, which are less
opaque, are sensitively traced by this line. Since most of the material in
a cloud resides in the envelopes, this line is adequate to estimate a
reliable mass for the cloud. To detect high-density clumps or cores inside 
clouds, with densities in the range 10$^4$ to $10^6$ \c3, other species like 
CS, C$^{18}$O or H$_2$CO (Mangum \& Wootten 1993, Bronfman, Nyman \& May 1996) 
or higher CO transitions (eg. Frail \& Mitchell 1998) need to be observed. 
Frail \& Mitchell (1998) estimated the densities of post-shock gas associated 
with OH 1720 MHz masers to be n$_{H_2} \sim 10^4 -  10^6$ \c3, and infer
pre-shock densities of the order $10^2$ to 10$^3$ \c3. In summary, our data 
reveal the ISM structures with which SNRs are interacting, but do not probe 
the shocked molecular gas in which the OH 1720 MHz maser emission is produced, 
which should be observed in other lines. 

\medskip
\subsection {Interaction between SNRs and molecular clouds}

Large molecular complexes are proposed to be the birthplace of the young,
massive progenitor stars of Type II supernovae. Half of the galactic SNRs
are supposed to be of this type, and this would be the reason why Huang \&
Thaddeus (1986) found a very good correlation between large molecular cloud
complexes and half of the SNRs surveyed in CO. It is likely that the observed
molecular complexes be part of the natal-clouds of the pre-supernova stars.
However, although it is statistically improbable that such a good correlation
arises from chance superpositions, there is very little direct evidence that
SNRs are actually in physical contact with the molecular clouds observed.

Shocked-excited 1720 MHz OH masers toward SNRs constitute a powerful tool to 
ensure on safe grounds that the molecular structures at the same velocity of the
masers are physically related to the remnants. The use of OH 1720 MHz masers as
signposts of SNR-molecular cloud interactions allowed us to detect the molecular
gas into which SN shock fronts are being driven. In what follows, we analize
the interaction of each SNR with its surrounding molecular gas.
\medskip
\subsubsection {G349.7+0.2}
In this case, the evidence for SNR-molecular cloud interaction is compelling,
based on the positional and velocity coincidence between the CO feature and the
OH 1720 MHz masers. This conclusion supports the statement of Clark \& Caswell 
(1976) that the expansion has been influenced by a dense environment in early 
stages. To obtain an additional proof of interaction, we have searched for IRAS 
point sources that fulfill the following conditions: (1) $S_{100}\geq 20$ Jy, 
(2) $1.2 \leq S_{100}/S_{60}\leq 6.0$, and (3) $S_{60}/S_{25}\geq 1$, where 
$S_\lambda$ denotes the (uncorrected) IR-flux density in the wavelength 
$\lambda \, \mu$m. These selection criteria are indicative of proto-stars 
(Junkes, F\"urst \& Reich 1992) or dust heated in SNR shocks (Arendt 1989). We 
have found that the point source IRAS 17146-3723, which agrees with these 
conditions, is positionally coincident with the CO cloud. It is unlikely that 
this IRAS source is a protostar triggered by the shock wave of G349.7+0.2 
because of the time scales involved. Formation of protostars from condensed 
molecular matter needs a period of the order of $10^6$ years, enough for the 
matter to cool, which is too long compared to the age of a SNR. We therefore 
suggest that IRAS 17146-3723 is produced by shock-heated dust and, together 
with the OH 1720 MHz maser emission, confirms the interaction between the SNR 
and the molecular cloud detected here. A thorough analysis of the more extended
region surveyed here, is given in Reynoso \& Mangum (2001).

\medskip
\subsubsection {CTB 37A}

Our observations show two distinct clouds towards CTB 37A, both coincident
spatially and in velocity with the two groups of OH 1720 MHz masers. Thus,
the present CO data provide additional support to the theory of the two 
overlapped SNRs. However, the complexity of the CO distribution in this 
direction makes it difficult to identify molecular clouds interacting with 
the SNR shock fronts. One of the remnants, G348.5+0.1, has a characteristic 
``breakout'' morphology, which would imply a strong density gradient. 
Nevertheless, there is no correspondence between the CO distribution and such 
radio morphology. This is in contrast with the case of 3C 391, where the 
$^{12}$CO distribution shows a molecular cloud exactly parallel to the bright 
radio emission ridge (Wilner et al. 1998). As in the case of G349.7+0.2, an
additional evidence for interaction is given by the IRAS point source 
17111-3824, which has an IR spectrum typical of shocked heated dust according
to the criteria of Junkes et al. (1992), and lies close to the OH maser at the 
northern cloud. This IRAS source is shown as a white star in Fig. 5.

Apart from the presence of the OH 1720 MHz masers and IRAS 17111-3824, there 
are no further indications of shocked molecular gas such as line broadenings or 
wings. However, at $\sim -88$ \k \ there appears a bright, elongated CO 
structure positionally following the masers closest to the central cloud. In 
Fig. 8, two profiles are shown, the upper towards the three masers at the 
central cloud (white crosses in Fig.  5), and the lower towards an arbitrary 
direction displaced from the remnant, at the northwest of the field. In the 
upper profile, three peaks are cleary distinguished: one at $\sim -65$ \k , 
which corresponds to the gas in which the OH 1720 MHz masers originate, and the 
others at approximately --88 and --107 \k. The peak at $\sim -107$ \k \ is also 
seen in the profile outside the SNR, and is commonly regarded as an expanding 
arm about 3 kpc from the Galactic center (Caswell et al. 1975). The other peak, 
instead, does not appear in the lower profile, so it is likely to be related 
with the central cloud. At --88 \k, the only CO emission observed extends over 
the area enclosed by the radio continuum emission. Therefore, we propose that 
the CO emission at this velocity corresponds to a fraction of molecular gas 
accelerated by the shock front towards the rear side of the remnant at a 
velocity of about 25 \k. This accelerated component of the gas would be 
additional proof for an SNR-molecular cloud interaction. 
 
Chevalier (1999) proposes that in the initial interaction, the (radiative) SNR 
shell drives a slab into the cloud. We suggest that in this case, the slab can
be seen as the --88 \k \ component. The post-shock pressure is $\rho_{cl} 
v^2_{cl} \simeq 8\times 10^{-9}$ dyn cm$^{-2}$, where the subindex $cl$ 
indicates that the parameters are estimated at the cloud. Assuming constant 
density for the radiative shell, and allowing for magnetic fields, Chevalier's 
model states that the ratio between the ram pressure in the slab and that at 
the front of the radiative shell, ${\rho_{cl} v^2_{cl}}/ {\rho_{0} v^2_{rs}}$, 
is limited to the range $\sim 10-100$. In this equation, $\rho_{0}$ is the 
intercloud density and $v_{rs}$ is the velocity of the radiative shell before 
entering the cloud. Thus, assuming $n_0=15$ \c3 \ as a typical value for the 
intercloud density (e.g. Chevalier 1999), the shock velocity at the cloud is 
$\sim 15-30$ \k. At velocities lower than $\sim 50$ \k, shocks are of C-type 
(Neufeld \& Dalgarno 1989), thus this estimate is consistent with the 
theoretical prediction by Lockett et al. (1999) that the conditions under which 
1720 MHz OH maser emission is produced can exist only if shocks are of C-type. 
An upper limit of $v_{rs}=60$ \k \ can be reached if an intercloud density as 
low as $n_0=1$ \c3 \ (e.g. McKee \& Cowie 1975, Dubner et al. 1999) is assumed.

\medskip
\subsubsection {G16.7+0.1}

Like G349.7+0.2, this is another case in which the evidence for SNR-molecular
cloud interaction is clear. What is remarkable about this case is that one
single maser was found to be associated with a region with several structures 
of molecular gas and more than one site of possible interaction between this 
gas and the SNR shock front. The same occurs with IC 443 which, in spite of 
being one of the remnants more clearly interacting with dense molecular gas, 
has only one maser coincident with a site where the shock is transverse 
(Claussen et al. 1997). 

Based on high-resolution observations, Frail \& Moffett (1993) describe the
central jetlike radio  structure as consisting of a center of symmetry at 
RA=$18^h 20^m 57^s.48$, decl=$-14^\circ 19^\prime 57\pp.6$ (J2000) from
which emission extended in two directions opens up into more extended lobes.
In Fig. 7, it can readily be seen that the northern lobe seems to be
inclined towards the central cloud, tending to enclose the inner CO 
emission peak (see also Fig. 6). Also, the southern lobe appears to the east of 
the southern cloud, with the OH 1720 MHz maser lying at the contact region 
between them. Both lobes show enhancements in the radio continuum emission at 
the locations of the molecular clouds, which is an indication of interaction. 

In virtue of the morphological agreement, the north-western cloud appears as an 
excellent candidate for interaction with this SNR, although no line 
broadenings are detected. The interaction between G16.7+0.1 and the northern
cloud places this remnant in the same class as IC 443 and 3C 391 in the sense 
that, although the morphologically most convincing site of SNR-molecular gas 
interaction is suggested by a flat radio continuum rim exactly parallel to the 
boundary of a molecular cloud, there appears only one maser (actually two in 
the case of 3C 391), and not at that site but near the opposite side of the 
remnant. Moreover, the morphology involved in such cases makes it difficult not 
to conceive a strongly transverse shock. Thus, it appears that not all 
transversely-shocked dense gas regions lead to the conditions which give rise 
to OH 1720 MHz maser emission. Possible explanations for the lack of OH 1720 MHz
masers in these cases could be related, for instance, to dissociation of OH by 
high velocity shocks. The trend observed in these three remnants is 
undoubtfully a topic that deserves further investigation,and should certainly 
be taken into account in future theoretical works.

\medskip
\section {Conclusions:}

The surroundings of the SNRs G349.7+0.2, CTB 37A and G16.7+0.1 have been 
investigated in the CO J=1-0 transition. These remnants have been selected
since all of them have associated OH 1720 MHz maser emission. Our observations
allowed us to determine the density distribution of the ISM into which these 
remnants are expanding, revealing a rich environment in all three cases. In 
spite of the confusion produced by the ubiquity of the CO J=1--0 emission, we 
have been able to identify molecular clouds interacting with SNRs based mainly 
on the positional and kinematical coincidence with shock-excited OH 1720 MHz 
masers, but also on morphological arguments as well as radio continuum 
enhancements, IR emission by shock-heated dust, and peculiar spectral 
components. Unfortunately, we have not detected spectral line broadenings, 
which are unmistakable signatures of interaction. However, there are other 
cases of SNRs interacting with molecular clouds, in which no line broadenings 
have been observed in the CO J=1--0 transition (Wilner et al. 1998, Dubner et 
al. 1999). Reasons for this non-detection vary from rapid dissociation of the 
molecules, masking of weak CO J=1--0 features by background cloud emission, and 
beam dilution of highly localized blast-wave interactions (Wilner et al. 1998).

The main results of this paper are summarized below.

{1)} All but two of the OH 1720 MHz masers are associated with CO emission 
peaks that posses an excellent velocity correlation with the former. The 
difference between both velocities is in average $(0.6\pm 1.7)$ \k \ in the
worst case. 

{2)} The masers are positionally close to, but not coincident with, CO peaks. 
The same trend has been observed by Frail \& Mitchell (1998).

{3)} The CO emission associated with the OH 1720 MHz masers allows constraint
of the systemic velocities of the SNRs.

{4)} The H$_2$ densities inferred ($\leq 10^3$ \c3) are representative of the
preshock gas, but do not trace high density clumps or shocked, compressed
material.

{5)} Several cases of SNR-molecular cloud interaction have been detected
based on wings, shifted spectral components, morphological coincidences and IR
emission from shock-heated dust.

{6)} In G16.7+0.1, the OH 1720 MHz maser seems to originate in molecular
gas accelerated to 6.5 \k , and not in the gas located at the systemic velocity 
of the SNR.

{7)} The interstellar gas does not account for the breakout morphology of
G348.5+0.1, one of the SNRs of the CTB 37A complex.

{8)} In CTB 37A, the shock is computed to be of C-type, in agreement with 
theoretical predictions (Lockett et al. 1999).

{9)} G16.7+0.1 is found to belong to a group of SNRs (together with 3C 391
and IC 443) with the following common characteristic: part of the rim is
obviously flattened by a dense molecular cloud. The morphology strongly
suggests that the expansion is perpendicular to the line of sight at this 
location. However, no OH maser emission at 1720 MHz is detected. The only 
OH 1720 MHz masers observed in such remnants (only one or two) are far 
from the flattened part of the rim. This behavior seems to contradict the 
suggestion that OH 1720 MHz masers appear where shocks are transverse.

\bigskip

We are grateful to the NRAO for allocating time in the 12  Meter Telescope
for this project. We also thank D. Helfand, N. Kassim and C. Salter for 
providing us their continuum images of the SNRs. E.M.R. acknowledges travel 
grant 1526/98 from CONICET (Argentina) for visiting the 12 Meter Telecope, as 
well as NRAO support and staff assistence during her visit.
This research was partially funded through a Cooperative Science Program
between CONICET and NSF and through CONICET grant 4203/96.

\section{Figure captions}

\figcaption[Figure1.ps]{A radio continuum image of G349.7+0.2 at 1.4 GHz 
obtained with the VLA, kindly provided by C. Salter. The grayscale is indicated 
in mJy/beam on top of the image, while the contours are expressed as a 
percentage of the peak flux density (0.48 Jy/beam) at 1, 2, 3, 5, 10, 20, 30, 
40, 50, 60, 70, 80, 90, and 99\%. For better contrast, white contours are used 
over dark shadows.  The white crosses show the positions of the OH 1720 MHz
masers detected by Frail et al. (1996). The beam, indicated in the bottom left 
corner, is 19$\pp.4 \times 4\pp.5$, P.A.=$-22^\circ.2$.}

\figcaption[Figure2.ps]{CO emission integrated between +14.5 and +18.5 \k \
toward G349.7+0.2, in greyscale and contours. Data were taken with the 12 Meter 
Telescope of the NRAO. The greyscale is in K \k , and is shown on top of
the image. The white contours represent the CO emission between 8 and 28 K 
\k , in steps of 4 K \k . The crosses show the positions of the OH 1720 MHz
masers.  The beam,  60$\pp \times 60\pp$, is indicated by the open circle in 
the top left corner. The 3$\sigma$ noise level is 6 K \k . A few radio 
continuum contours in grey are included to represent the SNR.} 

\figcaption[Figure3.ps]{A radio continuum image of CTB 37A at 1.4 GHz, from 
Kassim et al. (1991), obtained with the VLA. Two sources can be identified: 
G348.7+0.1 to the west, and G348.5-0.0 to the east. The grayscale is indicated 
in mJy/beam on top of the image, while the contours are at 1, 3, 5, 10, 20, 30, 
65, and 95 \% of the peak flux. Here again, white contours are used over dark 
shadows. The peak flux is 0.43 Jy/beam. Black crosses show the positions of the 
OH 1720 MHz masers between --63 and --67 \k , while white crosses represent the
two masers at --25.3 and --22.7 \k \ (Frail et al. 1996). The beam, indicated 
in the bottom left corner, is 32$\pp.9 \times 18\pp$, P.A.=$4^\circ.8$.}

\figcaption[Figure4.ps]{CO emission integrated between --25.3 and --22.7 \k \
toward CTB 37A. Data were taken with the 12 Meter Telescope of the NRAO.
The greyscale is in K \k , and is shown on top of the image. The beam, 
53$\pp.6 \times 53\pp.6$, is plotted in the bottom left corner. A few 
representative contours of the continuum emission from CTB 37A are included as 
grey lines. CO emission contours at 11.5, 17, 22 and 28.5 K \k \ are
plotted as black lines, except for the last one which appears in white. The 
two white plus signs represent the two OH 1720 MHz masers at --21.4 and --23.3 
\k \ (Frail et al. 1996). The 3$\sigma$ noise level is 9 K \k.}

\figcaption[Figure5.ps]{CO emission integrated between --68.3 and --60.4 \k \
toward CTB 37A. Data were taken with the 12 Meter Telescope of the NRAO.
The greyscale is in K \k , and is shown on top of the image. The beam size
is 60$\pp \times 60\pp$. A few representative contours of the continuum 
emission from CTB 37A are included as thick grey lines. CO emission contours 
at 20, 30, 40, 55, 75, and 95 \% of the peak emission are plotted in black or 
white solid lines, according to the darkness of the background. The peak 
emission is 82.2 K \k . The plus signs show the positions of the OH 1720 MHz
masers found by Frail et al. (1996) between --63 and --67 \k , where white
lines have been used over dark background for better contrast. The white
star represents the IRAS point source 17111-3824 discussed in the text (Section 
4.2.2). The arrows indicate the three clouds discussed in the text. The 
3$\sigma$ noise level is 7 K \k.}

\figcaption[Figure6.ps]{A radio continuum image of G16.7+0.1 at 4.86 GHz from 
Helfand et al. (1989), obtained with the VLA. The grayscale is indicated in 
mJy/beam on top of the image, while the contours are at 1, 5, 15, and 30 
mJy/beam. The plus sign indicates the OH 1720 MHz maser detected by Green et
al. (1997). The beam, indicated in the bottom left corner, is 21$\pp \times 
16\pp$, P.A.=$11^\circ$.}

\figcaption[Figure7.ps]{CO emission integrated between --25.1 and --25.9 \k \
toward G16.7+0.1. Data were taken with the 12 Meter Telescope of the NRAO.
The greyscale is in K \k , and is shown on top of the image. The beam size,
53$\pp.6 \times 53\pp.6$, is plotted in the bottom right corner. A few 
representative contours of the continuum emission from G16.7+0.1 are included 
as dark-grey lines. CO emission contours at 3.8, 7.6, 12.7, 16.5, 20.3, and 24
K \k \ are plotted in black and white solid lines, according to the darkness of 
the background. The plus sign in white shows the position of the OH 1720 MHz
maser found by Green et al. (1997). The arrows indicate the three clouds 
discussed in the text. The 3$\sigma$ noise level is 1.5 K \k.} 

\figcaption[Figure8.ps]{CO emission profiles toward the OH masers lying on the
central cloud in CTB 37A (upper profile) and towards an arbitrary direction, to 
the northwest of the observed field and displaced from the continuum emission 
(lower profile). Brightness temperatures are in Kelvins, while velocities are 
in \k .}


\begin{thebibliography}{DUM}
 
\bibitem[1973]{}Allen, C.W. 1973, Astrophysical Quantities (3rd ed.; London: Athlone)
\bibitem[1989]{}Arendt, R. 1989, ApJS 70, 181
\bibitem[2000]{}Brogan, C. L., Frail, D. A., Goss, W. M., \& Troland, T. H.:
2000, ApJ 537, 875
\bibitem[1996]{}Bronfman, L., Nyman, L.-\AA., \& May, J.  1996, A\&AS 115, 81
\bibitem[1998]{}Burton, M.G., Brand, P.W.J.L., Geballe, T.R., Webster, A.S. 
1988, MNRAS 231, 617
\bibitem[1998]{}Case, G.L., \& Bhattacharya, D. 1998, ApJ 504, 761
\bibitem[1975]{}Caswell, J.L., Murray, J.D., Roger, R.S., Cole, D.J., \&
Cooke, D.J. 1975, A\&A 45, 239
\bibitem[1999]{}Chevalier, R.A. 1999, ApJ 511, 798
\bibitem[1976]{}Clark, D.H. \& Caswell, J.L. 1976, MNRAS 174, 267
\bibitem[1997]{}Claussen, M.J., Frail, D.A., Goss, W.M., \& Gaume, R.A. 1997, 
ApJ 489, 143
\bibitem[1979]{}DeNoyer, L.K. 1979a, ApJ 228, L41
\bibitem[1979]{}DeNoyer, L.K. 1979b, ApJ 232, L165
\bibitem[1979]{}DeNoyer, L.K. 1983, ApJ 264, 141
\bibitem[1990]{}Digel, S.W., Bally, J., \& Thaddeus, P. 1990, ApJ 357, L29
\bibitem[1997]{}Digel, S.W., Hunter, S.D., Mukherjee, R., De Geus, E.J.,
Grenier, I.A., Heithausen, A., Kanbach, G., \& Thaddeus, P. 1997, in IAU
Symp. 170, CO: Twenty-Five Years of Millimeter-Wave Spectroscopy, ed. W.B.
Latter et al. (Dordretch: Kluwer), 22
\bibitem[1999]{}Dubner, G.M., Giacani, E. B., Reynoso, E. M., Goss, W.  M., 
Roth, M., Green, A.J.  1999, AJ 118, 930
\bibitem[1976]{}Elitzur, M. 1976, ApJ 203, 124
\bibitem[1989]{}Fich, M., Blitz, L., \& Stark, A.A. 1989, ApJ 324, 272
\bibitem[1996]{}Frail, D.A., Goss, W.M., Reynoso, E.M., Giacani, E.B., Green, 
A.J., \& Otrupcek, R. 1996, AJ 111,1651
\bibitem[1994]{}Frail, D.A., Goss, W.M., \& Slysh, V.I. 1994, ApJ 424, L111
\bibitem[1998]{}Frail, D.A., \& Mitchell, G.F. 1998, ApJ 508, 690
\bibitem[1993]{}Frail, D.A., \& Moffett, D.A. 1993, ApJ 408, 637
\bibitem[1968]{}Goss, W.M. 1968, ApJS 15, 131
\bibitem[1968]{}Goss, W.M., \& Robinson, B.J. 1968, Astrophys. Lett. 2, 81
\bibitem[1997]{}Green, A.J., Frail, D.A., Goss, W.M., \& Otrupcek, R. 1997, AJ 
114, 2058
\bibitem[1971]{}Hardebeck, E.G. 1971, ApJ 170,281
\bibitem[1989]{}Helfand, D.J., Velusamy, T., Becker, R.H., \& Lockman, F.J. 
1989, ApJ 341, 151
\bibitem[1986]{}Huang, Y.-L., \& Thaddeus, P. 1986, ApJ 309, 804
\bibitem[1992]{}Junkes, N., F\"urst, E., \& Reich, W. 1992, A\&A 261, 289
\bibitem[1991]{}Kassim, N.E., Baum, S.A., \& Weiler, K.W. 1991, ApJ 374, 212
\bibitem[1998]{}Koralesky, B., Frail, D.A., Goss, W.M., Claussen, M.J., \& 
Green, A.J. 1998, AJ 116, 1323
\bibitem[1989]{}Landecker, T.L., Pineault, S., Routledge, D., \&  Vaneldik, 
J.F. 1989, MNRAS 237, 277
\bibitem[1983]{}Lebrun, F., Bennett, K., Bignami, G. F., Caraveo, P. A., 
Bloemen, J. B. G. M., Hermsen, W., Buccheri, R., Gottwald, M., Kanbach, G., 
Mayer-Hasselwander, H. A. 1983, ApJ 274, 231
\bibitem[1999]{}Lockett, P., Gauthier, E., \& Elitzur, M. 1999, ApJ 511, 235
\bibitem[1993]{}Mangum, J.G., \& Wootten, A. 1993, ApJS 89, 123
\bibitem[1975]{}McKee, C.F., \& Cowie, L.L. 1975, ApJ 195, 715
\bibitem[1988]{}Mead, K.N., \& Kutner, M.L. 1988, ApJ 330, 399
\bibitem[1979]{}Milne, D.K., Goss, W.M., Haynes, R.F., Wellington, K.J., \&
Caswell, J.L. 1979, MNRAS 188, 437
\bibitem[1989]{}Neufeld, D.A., \& Dalgarno, A. 1989, ApJ 340, 869
\bibitem[1998]{}Oka, T., Hasegawa, T., Hayashi, M., Handa, T., \& Sakamoto,
S. 1998, ApJ 493, 730
\bibitem[1999]{}Reach, W.T., \& Rho, J. 1999, ApJ 511, 836
\bibitem[1995]{}Reynoso, E.M., Dubner, G.M., Goss, W.M., \& Arnal, E.M. 1995, 
AJ 110, 318
\bibitem[2001]{}Reynoso, E.M., \& Mangum, J.G. 2001, AJ, in press
\bibitem[1998]{}Seta, M., Hasegawa, T., Dame, T.M., Sakamoto, S., Oka, T., 
Handa, T., Hayashi, M., Morino, J.-I., Sorai, K., \& Usuda, K.S.  1998, ApJ 
505, 286
\bibitem[1969]{}Turner, B.E. 1969, ApJ 157, 103
\bibitem[1987]{}White, G.J., Rainey, R., Hayashi, S.S., \& Kaifu, N. 1987, A\&A 
173, 337 
\bibitem[1997]{}Wilner, D.J., Reynlolds, S.P., \& Moffett, D.A. 1998, AJ 115, 
247
\bibitem[1981]{}Wootten, A. 1977, ApJ 216, 440
\bibitem[1981]{}Wootten, A. 1981, ApJ 245, 105
\end{thebibliography}
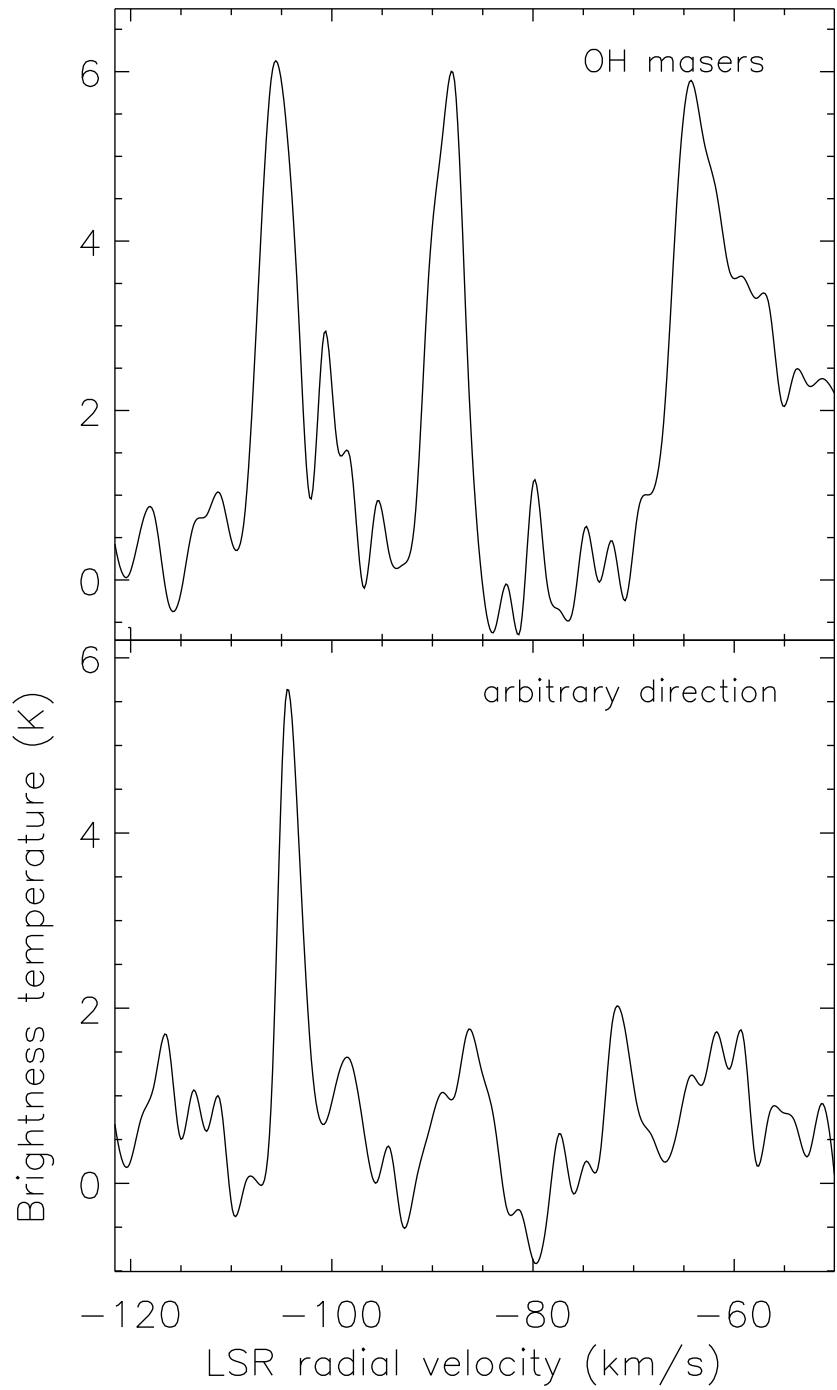
\end{document}